\documentclass[12pt]{article}

\usepackage{amsmath,amssymb}
\usepackage{graphicx}
\usepackage{dcolumn}
\usepackage{bm}
\usepackage[numbers,super,comma,sort&compress]{natbib}
\usepackage{multirow}
\usepackage{caption}
\usepackage{subcaption}

\begin{document}

  \makeatletter
  \renewcommand\@biblabel[1]{#1.}
  \makeatother

\bibliographystyle{apsrev}

\renewcommand{\baselinestretch}{1.5}
\normalsize


{\sffamily
\noindent {\bfseries \Large Polyelectrolyte Polypeptide Scaling Laws Via Mechanical and Dielectric Relaxation Measurements}\\

\noindent {Jorge Monreal\\}

\vspace{-8mm}

\noindent {University of South Florida}\\
}

\vspace{-5mm}

\noindent {\em Dated: \today}\\

\begin{center}
\begin{minipage}{5.5in}

{\bf ABSTRACT:}  
Experimental results from mechanical viscoelastic as well as dielectric relaxation times were compared to theoretical expectations utilizing polymer scaling theory.  Viscoelastic relaxation of a hydrogel at 33\% relative humidity fabricated from co-poly-L-(glutamic acid$_{4}$, tyrosine$_{1}$) [PLEY(4:1)] crosslinked with poly-L-lysine  scaled with concentration according to reptation dynamics.  High frequency dielectric relaxation of aqueous copolymer PLEY(4:1) scaled with concentration as an ideal chain and aqueous Poly-L-glutamic acid scaled as an extended chain.  The study shows that two seemingly different measurement methods can yield information about the state of polymer chain conformation \emph{in situ}. 


\end{minipage}
\end{center}

\clearpage

\section*{\sffamily \large INTRODUCTION}
Currently, there has been wide interest in utilizing aqueous polypeptide solutions to electrospin nanofibers for various types of medical applications.\cite{kh12}  Recent studies have electrospun several different synthetic polypeptides from water.\cite{kh11}  One of the tenets of efficient electrospinning is that polymer chains must be highly overlapped.\cite{shenoy05}  Typical nanofiber research with aqueous synthetic polypeptide melts relies on concentrations near the solubility limit to ensure chain overlap.  However, high concentration does not necessarily dictate spinnability.  The analysis method presented here could provide insight into the spinnability of a synthetic polypeptide polymer melt drawn from chain conformation.  

Here we have studied polyelectrolyte polypeptides (PEPs), which are polypeptides with electrolyte repeating groups.  Polyelectrolyte solutions have solution properties unlike that of neutral polymers.\cite{strobl,dob95}  In stark contrast to most neutral polymers PEPs are soluble in water because of amino acid side chain ionization.  Ions remain on the side chain while counter-ions surround them in solution.   However, counter-ions do not diffuse away to be distributed homogeneously within the solvent.  Rather, strong attractive forces between the cation (anion) and counter-ion prevent diffusive motility.  Some of the counter-ions are kept within the neighborhood of the polyion as condensates so the effective charge density is reduced.  Only at distances larger than the Bjerrum length, $\xi_B$, do charges appear to have mobility.  Presence of oppositely charged counter-ions screens polyionic potential.  For a charge in the immediate vicinity, $r$, of an ion the Coulomb attraction is quite strong.  Coulombic potential decreases at increasing,  $r$,  away from the polyion as $\sim e^{-r/\xi_D}$ and finally vanishes for $r \ge \xi_D$.  Here, the Debye length, $\xi_D$,  effectively describes the size of a screening charge cloud surrounding an ion and is an inverse function of ionic strength.  Increased ionic strength decreases, $\xi_D$, and screening sets in at decreased distances from the ion.  Therefore, increasing the PEP concentration increases counterion concentration, which in turn increases the ionic strength. As a result, $\xi_D$ decreases.  A decreased Debye length results in larger charge screening. Screening can be so strong that PEPs lose their unique properties and start behaving as neutral polymer chains \cite{strobl}.  This happens when the Debye length drops below the Bjerrum length, $\xi_D < \xi_B$.  As this study was conducted with high concentrations close to solubility limits, polymers turned out to behave as as neutral polypeptides.   

\begin{figure} [ht]
\centering
\begin{subfigure}[b]{.5\textwidth}
  \centering
  \graphicspath{ {./DisPics/} }
  \includegraphics[width=.5\linewidth]{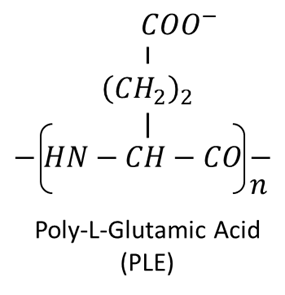}
  \caption {}
\end{subfigure}%
\begin{subfigure}[b]{.5\textwidth}
  \centering
  \graphicspath{ {./DisPics/} }
  \includegraphics[width=.7\linewidth]{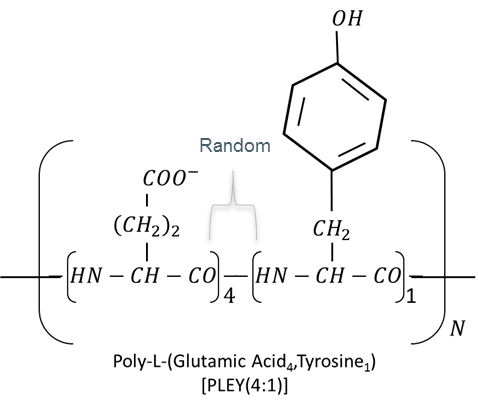}
  \caption {}
\end{subfigure}
\caption{\emph{Chemical structures of a.) PLE.;  b.) PLEY(4:1). }}
\label{pls}
\end{figure}

Figures \ref{pls}a,b show the polyelectrolyte structures of Poly-(L-Glutamic Acid) [PLE] and Poly-L-(Glutamic Acid$_4$, Tyrosine$_1$) [PLEY(4:1)], respectively.  The glutamic acid residue is a weak polyacid in solution.  The tyrosine amino acid remains neutral but is a polar molecule with hydrophobic tendencies. 

From general polymer basics it is known that $\Theta$-solvents allow for a balance of interaction forces between polymer chain and solvent.  Thus, a polymer settles into an an ideal, non-perturbed chain configuration.  In good solvent, chain monomers try to maximize interactions with the solvent and the chain becomes expanded.  Therefore, the polymer is found in different configurations depending on the solvent.  

Polymers are known to be fractal objects.\cite{degn,strobl,doi}   Down to a limit, dimensionality of the polymer varies with scale due to self-similarity.  Theory due to Flory established a scaling exponent given by $\nu = 1/d_f$, where $d_f$ is a fractal dimension.\cite{flory1}  For an ideal chain with random walk, $d_f=2$, therefore, $\nu=1/2$.  Thus, a theta solvent has $\nu=1/2$ and a good solvents has $\nu=3/5$.  It has been found that a more accurate number is $\nu=0.588$.

With regards to scaling of relaxation times as a function of polymer concentration, Rubinstein and Colby present the following for neutral polymers: \cite{rub12}
\begin{eqnarray}
\textit{ ($\nu=0.588$)               }&:& \ \tau \sim N^2 c^{(2-3\nu)/(3\nu-1)} \ \to \tau_{gs} \sim N^2 c^{0.31} \label{up} \\
\textit{ ($\nu=1/2$)               }&:& \ \tau \sim N^2 c^{(2-3\nu)/(3\nu-1)} \ \to \tau_{\Theta s} \sim N^2 c \label{tup} \\
\textit{Reptation ( $\nu=0.588$)               }&:& \ \tau_d \sim N^3 c^{3(1-\nu)/(3\nu-1)} \ \to \tau_{dgs} \sim N^3 c^{1.6} \label{repent}.
\end{eqnarray}

Equations \ref{up}--\ref{repent}, respectively, give relaxation times as: $\tau_{gs}$ for unentangled neutral polymers in a good solvent; $\tau_{\Theta s}$ for unentangled neutral polymers in a $\Theta$-solvent; and polymer relaxation according to reptation dynamics $\tau_{dgs}$ for entangled neutral polymers in a good solvent.  These differ significantly from scaling laws for semi-dilute, polyelectrolyte solutions which are given as : \cite {rub12, bordi00, bordi99, bordi04}
\begin{eqnarray}
\textit{Unentangled                }&:& \tau_{pe} \sim N^2 c^{-1/2} \label{upe} \\
\textit{Entangled                }&:& \tau_{dpe} \sim N^3 c^{0} \label{reppe}.
\end{eqnarray}
It will be seen that the samples in this study, though polyectrolyte polypeptides, behaved as given by equations \ref{up}, \ref{tup} and \ref{repent}.  That was due to the high concentrations involved.

\begin{figure} [ht]
\centering
\begin{subfigure}[b]{.5\textwidth}
  \centering
  \graphicspath{ {./DisPics/} }
  \includegraphics[width=.9\linewidth]{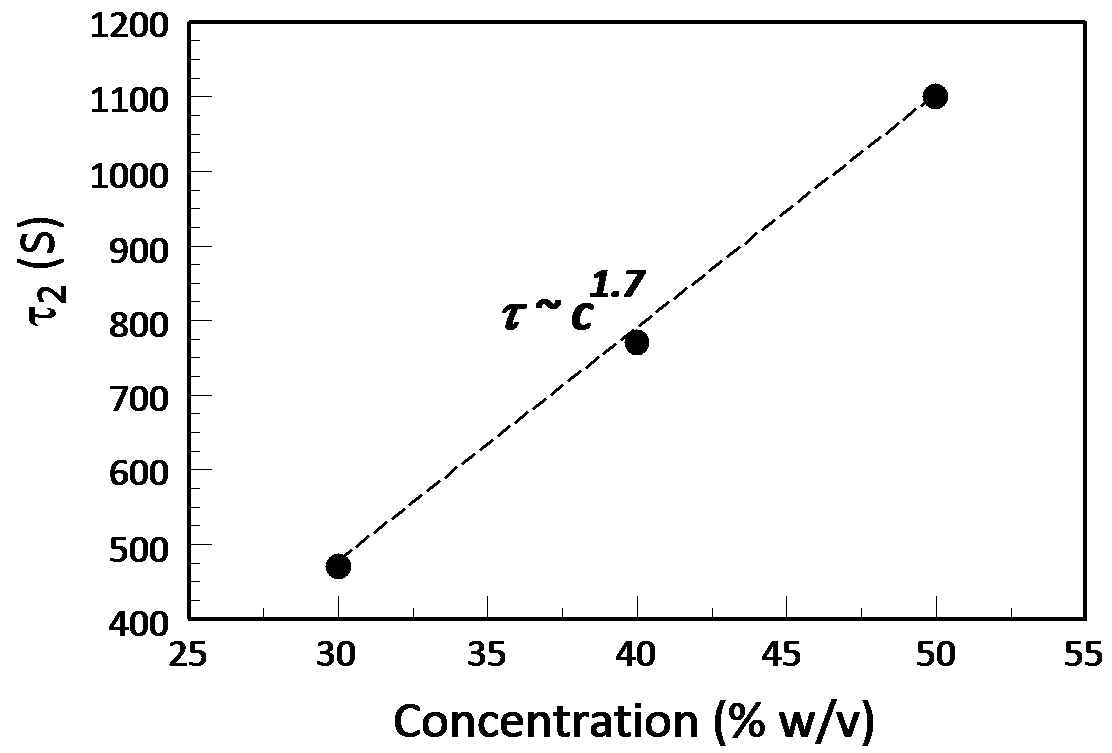}
  \caption {}
\end{subfigure}%
\begin{subfigure}[b]{.5\textwidth}
  \centering
  \graphicspath{ {./DisPics/} }
  \includegraphics[width=.9\linewidth]{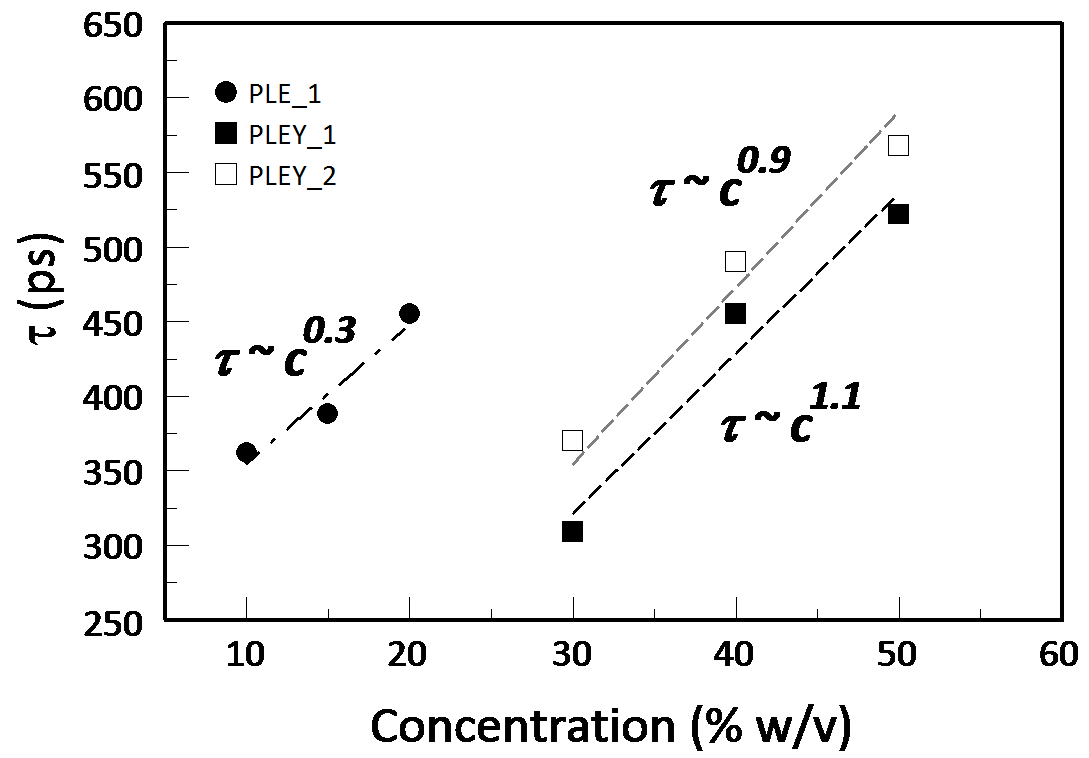}
  \caption {}
\end{subfigure}
\caption{\emph{Relaxation time versus concentration (\% w/v). a.) Relaxation time of PLEY(4:1) viscoelastic material versus concentration taken from Table \ref{table1} at three different strains.;  b.) Power law scaling of PLEY(4:1) and PLE.\cite{mon3}}}
\label{t2l}
\end{figure}

\begin{table}[ht]
  \centering
   \caption{\emph{Relaxation times for three representative PLEY(4:1) crosslinked hydrogel samples at 33\% relative humidity at high strain.(Adapted from Monreal et al.).\cite{mon2}}}
    \begin{tabular}{|r|c|}
    \hline
    \multicolumn{2}{|c|}{High Strain}  \\ 
    \multicolumn{2}{|c|}{($\varepsilon$ = 0.10)}  \\ \hline
   \multicolumn{1}{|c|}{Sample: Nominal concentration (\%w/v)}   &  $\tau_2$ (s)   \\ \hline
    \multicolumn{1}{|c|}{310233-1: 30} & 470    \\ \hline
    \multicolumn{1}{|c|}{415433-2: 40} & 770    \\ \hline
    \multicolumn{1}{|c|}{520333-1: 50}  & 1100  \\ 
   \hline
   \end{tabular}
  \label{table1}
\end{table}
Previous studies showed that a hydrogel of poly-L-(glutamic acid$_{4}$, tyrosine$_{1}$) crosslinked with poly-L-lysine (PLL) is highly viscoelastic at 33\% relative humidity.\cite{mon2}  Table \ref{table1} presents the relaxation times at high strain for samples at three different PLEY nominal concentrations.  Data is taken from Monreal et al.\cite{mon2} In that study, PLEY(4:1) was crosslinked with PLL via a carbodiimide crosslinker.  The hydrogel was allowed to reach equilibrium at 33\% relative humidity.  Samples were subjected to low, medium and high strains and relaxation times recorded.  It was found that crosslinked polymer network bore the full brunt of the load at the highest strain.  It would be expected that the force applied to the crosslinked polymer network leads to chain relaxation according to reptation dynamics.  Indeed such relaxation was found through the study here.  Strain-induced relaxation studies of PLEY(4:1) viscoelastic material showed that relaxation at high strain scales as an entangled neutral polymer in good solvent according to the reptation model of equation \ref{repent}.  Figure \ref{t2l}a shows that the high strain curve  scales close to $\tau \sim c^{1.6}$.  

Useful insights were also obtained from dielectric relaxation studies of PLE and PLEY(4:1) in aqueous solutions. Data for this study was obtained from Monreal et al.\cite{mon3}  There, dielectric relaxation of PLE and PLEY aqueous solutions were studied in the radio frequency range of 50 - 800 MHz with an impedance analyzer.  Dielectric relaxation peaks appeared at certain frequencies dependent on the PEP and its concentration.  The points shown in Fig. \ref{t2l}b represent peak relaxation times for each PEP and related concentration.  To study scaling behavior of the material, it was first necessary to estimate the critical overlap volume fraction using $\phi^{*} \sim N ^{-4/5}$.  For both PLE and PLEY we assume 50\% of glutamic acids are ionized at neutral pH and lyophilized in sodium salt. For PLEY in particular we assume tyrosine is neutral.  Haynie, thus, obtained for PLEY(4:1) of $ M_w \approx 30\ kDa$ [$ \rho \approx 1.54 \  g \ cm^{-3} ; N \approx 169$] and for PLE of $ M_w \approx 30\ kDa$ [$\rho \approx 1.52 \ g \ cm^{-3} ; N \approx 214$].\cite{dth15}.  As approximation we can estimate $N\approx200$ for both PLE and PLEY.  Thus, the critical overlap volume fraction $\phi^{*} \sim (200)^{-4/5} = 0.014$.  We can compare $\phi^{*}$ to experimental volume fraction via $\phi =c/\rho$, where $c$ is polymer concentration in w/ v ($g \ cm^{-3}$).  We, thus, obtain the following volume fractions:  PLEY ($\phi_{30\%} = 0.20 $; $\phi_{40\%} = 0.26 $; $\phi_{50\%} = 0.33$) which are about $\mathrm{14 \ to \ 23 \ times}\ \phi^{*}$; and PLE ($\phi_{10\%} = 0.07 $; $\phi_{15\%} = 0.10 $; $\phi_{20\%} = 0.13$) which are about $\mathrm{5 \ to \ 9 \ times}\ \phi^{*}$.  So we find that  $\phi_{PLEY} \gg  \phi^{*}$ and $\phi_{PLE} >  \phi^{*}$.  In general, both PLE and PLEY are in a regime where there is chain overlap.  PLEY is in the concentrated regime, but it is possible that the lower concentration of PLE could be in the semi-dilute regime and reaches the concentrated regime at 20\% w/v.  Regardless, Fig.\ref{t2l}b shows that PLE scales as a neutral polymer not a semi-dilute polyelectrolyte.  That is it scales with $\tau \sim c^{0.3}$ not $\tau \sim c^{0}$.

Figure \ref{t2l}b also shows that PLEY(4:1) scales as $\tau \sim c^{0.9-1.1}$, which is close to the scaling law predicted for neutral polymers in a $\Theta$-solvent.  Deviations appear due to experimental variations.  Nonetheless, PLEY(4:1) evidently behaves as an ideal chain with $\nu=1/2$!  This is expected for neutral polymers as de Gennes explains that for concentrated polymer melts each polymer chain experiences opposing forces throughout so that there are no net expansive forces.\cite{degn}  The result is polymer conformation as an ideal chain.  Additionally, the chemical structure of PLEY (Fig.\ref{pls}b) contains a hydrophilic amino acid residue (glutamic acid) that wants to maximize contact with the solvent.  Opposing that tendency is the hydrophobic tyrosine amino acid residue.  These two phenomena seem to result in PLEY behaving as an ideal chain at the concentrations studied. 

A surprising find was the scaling trend for PLE.  As shown in Fig.\ref{t2l}b, PLE scaled as $\tau \sim c^{0.3}$.  This scaling changes with concentration analogously to a neutral polymer in good solvent of equation \ref{up}!  The possibility that PLE solutions could have been in the dilute regime was discounted by the estimate that $\phi_{PLE} >  \phi^{*}$.  In good solution a polymer exhibits chain swelling due to excluded volume effects.  Given that PLE is a polyanion there seems be excluded volume effects possibly due to monomer-monomer charge repulsion.  As a consequence, PLE polymer chains might not entangle in a manner conducive to electrospinning even at higher concentrations.  A main reason could be due to chain stiffness.  Monomer charge repulsion results in stiffening of a polymer chain.  It has been found that polymer chain elasticity is an essential component for the fabrication of nano-fibers by electrospinning.\cite{yu06}    

Within the range of polymer concentrations studied, chains of aqueous PLEY(4:1) overlap in the same manner as highly concentrated neutral polymers.  Theory would expect this PEP to produce nano-fibers through electrospinning.  Experiments by Khadka et al confirm such expectation.\cite{kh11}  In contrast, in a different study Khadka et al. found that PLE did not spin under any conditions.\cite{kh14}.  Results from this scaling study might offer an explanation. The group also found that PLL was not spinnable, but poly-L-orthonine (PLO) did produce nano-fibers by electrospinning.\cite{kh11}  It would be interesting to see if relaxations times of PLO polymer chains scale as ideal chains with increasing concentration.

The most interesting, albeit expected, result is that neither PLEY nor PLE scaled as polyelectrolytes such as $\tau \sim c^{-0.5}$, for semi-dilute solutions or $\tau \sim c^0$ for concentrated entangled solutions.  Rather, both scaled as neutral polymers.

\clearpage


\bibliography{scalepoly}   

\clearpage

\clearpage


\end{document}